\documentclass{iopart}
\usepackage{graphicx}
\usepackage{color}
\usepackage{ulem}

\begin{document}

\title{\bf Influence of firing mechanisms on gain modulation}
\author{Ryota Kobayashi}
\address{Kyoto University Department of Physics,  Kyoto 606-8502, Japan}
\ead{kobayashi@ton.scphys.kyoto-u.ac.jp}

\begin{abstract}
  We studied the impact of a dynamical threshold on the {\it f-I} curve$-$
  the relationship between the input and the firing rate of a neuron$-$in the presence of background synaptic inputs. 
  First, we found that, while the leaky integrate-and-fire model cannot reproduce the {\it f-I} curve of a cortical neuron, the leaky integrate-and-fire model with dynamical  threshold can reproduce it very well.
  Second, we found that the dynamical threshold modulates the onset and the asymptotic behavior of the {\it f-I} curve.
   These results suggest that a cortical neuron has an adaptation mechanism and that the dynamical threshold has some significance for the computational properties of a neuron.
\end{abstract}
\maketitle

%%%%%%%%%%%%%%%%%%%%%%%%%%%%%%
%%%   Introduction
%%%%%%%%%%%%%%%%%%%%%%%%%%%%%%
\section{Introduction}
  Neurons receive thousands of synaptic inputs, which are then transformed into output spike trains.  A common assumption in neuroscience is that the firing rate$-$the average number of spikes per unit time$-$of a neuron conveys information and is used for computation in the brain.  This assumption is supported by a number of experimental studies~\cite{Adrian1926, Barlow1972, Newsome1989}.
  It is therefore very important to study the relationship between the input current and the firing rate ({\it f-I} curve) in the presence of background synaptic inputs in order to understand the computation performed by a neuron.\\
 Recently, the leaky integrate-and-fire with dynamical threshold (LIFDT) model~\cite{Chacron2001, Chacron2000, Geisler1966, Jolivet2008, Koch1999, LaCamera2004, Liu2001} has been proposed so as to include spike frequency adaptation. Although the LIFDT model is simple, it can reproduce important properties of cortical neurons such as the negative interspike interval correlation~\cite{Chacron2001, Liu2001} and the {\it f-I} curve of pyramidal neurons driven by a fluctuating current~\cite{LaCamera2004}. It is also reported that the spike threshold of a cortical neuron {\it in vivo} is not constant but depends on the preceding spike times~\cite{Henze2001}.\\
 Our understanding of the effect of the dynamical threshold on the computational properties of a neuron is lacking. We studied whether the leaky integrate-and-fire (LIF) and LIFDT neuron can reproduce the {\it f-I} curve of a cortical neuron~\cite{Chance} and how the dynamical threshold modulates the {\it f-I} curve in the presence of background synaptic inputs.

%%%%%%%%%%%%%%%%%%%%%%%%%%%%%%
%%%  Method
%%%%%%%%%%%%%%%%%%%%%%%%%%%%%%
\section{Method}
\subsection{Leaky integrate-and-fire neuron}
	Here, we briefly introduce the leaky integrate-and-fire (LIF) neuron~\cite{Koch1999}.  The membrane potential $V$ of the neuron obeys the first-order differential equation
\begin{eqnarray}
	&& C \frac{dV}{dt}= - g_{ {\scriptstyle \rm{L} } } (V-V_{\infty})+ I(t),  \label{eq:LI_eq} 
\end{eqnarray}
where $C$ is the membrane capacitance, $g_{\scriptstyle \rm{L} }$ is the leak conductance, $V_{\infty}$ is the resting potential of the neuron, and $I(t)$ is the input current.
When the membrane potential reaches the threshold $\theta_0$, a spike is generated and we instantaneously reset $V(t)$ to the resetting potential $V_{\rm r}$.  We adopted the parameters of the LIF neuron from Chance et al.~\cite{Chance}, where $C= 0.5$[$\mu$F/$\rm cm^2$], $g_{\scriptstyle \rm{L} }= 0.025$[$\mu$S/$\rm cm^2$], $V_{\infty}= -65$[mV], $V_{\rm r}= -60$[mV], and $\theta_0= -54$[mV].

\subsection{Leaky integrate-and-fire neuron with dynamical threshold }
 We briefly introduce here the LIFDT neuron~\cite{Chacron2001, Chacron2000,Geisler1966, Jolivet2008, Koch1999, LaCamera2004, Liu2001}.  The membrane potential $V$ of the neuron obeys Eq. (\ref{eq:LI_eq}).  In the following, we take the parameters $C$, $g_{\scriptstyle \rm{L} }$, $V_{\infty}$, $V_{\rm r}$ to be the same as in the LIF neuron.
When the membrane potential reaches the threshold $\theta(t)$, a spike is generated and we instantaneously reset $V(t)$ to the resetting potential $V_{\rm{r} }$.  The emission of a spike causes the threshold to increase by an amount $A_{\theta}$ and then decay to its resting value $\theta_{\infty}$ exponentially: 
\begin{eqnarray}
	&& \frac{d\theta}{dt}= -\frac{\theta-\theta_{\infty} }{\tau_{\theta} }+ A_{\theta} \sum_k \delta(t-t_k), \label{eq:Dyn_th}
\end{eqnarray}
where $\tau_{\theta}$ is the time constant of the dynamical threshold and $t_k$ is the $k$-th spike time. The sum is taken over all the spikes generated by the neuron up to time $t$. 
The initial condition of the threshold is its resting value: $\theta(0)= \theta_{\infty}$.  The parameters of the dynamical threshold are $\theta_{\infty}= -54$[mV], adopted from Chance et al.~\cite{Chance}., and $\tau_{\theta}= 80$[ms], adopted from Liu and Wang~\cite{Liu2001}.
\subsection{Synaptic input}
 We considered an input current $I(t)= m+ I_{\rm s}(t)$, which is the sum of a constant current $m$ and a background synaptic current $I_{\rm s}$.  The synaptic current is described by
\begin{eqnarray}
	&& I_{\rm s}(t)=  g_{\scriptstyle \rm{E} }(t) (V_{\scriptstyle \rm{E} }- V(t))+ g_{\scriptstyle \rm{I} }(t) (V_{\scriptstyle \rm{I} }-V(t)),  \label{eq:I_syn} \\
	&& g_{\scriptstyle \rm{E, I}}(t)= \sum_k a_{\scriptstyle \rm{E, I} } \exp \left( -\frac{t-t^{\scriptstyle \rm{E, I} }_k}{\tau_{\scriptstyle \rm{E, I} } } \right), \nonumber
\end{eqnarray}
where $g_{\scriptstyle  \rm{E, I}}$ are the excitatory (E) and the inhibitory (I) synaptic conductances,  $V_{\scriptstyle  \rm{E, I}}$ are their respective reversal potentials, $\tau_{\scriptstyle \rm{E,I} }$ are their respective time constants, $a_{\scriptstyle  \rm{E, I} }$ are their respective peak conductances, and $t^{\scriptstyle \rm{E, I} }_k$ are the $k$-th spike times of the respective presynaptic neuron.  The spike times of the respective presynaptic neuron are generated by an independent homogeneous Poisson process with the same rate $\gamma$.
The parameters are $\tau_{\scriptstyle \rm{E} }= 5$[ms], $\tau_{\scriptstyle \rm{I} }= 10$[ms], $V_{\scriptstyle \rm{E} }= 0$[mV], $V_{\scriptstyle \rm{I} }= -80$[mV], $a_{\scriptstyle \rm{E} }= 0.01$[$\mu$S/$\rm cm^2$], and $a_{\scriptstyle \rm{I} }= 0.04$[$\mu$S/$\rm cm^2$], adopted from Chance et al.~\cite{Chance}.  
\subsection{{\it f-I} curve}
For the LIF and LIFDT neurons, we calculated the {\it f-I} curves, which give the relationship between the constant current $m$ and the firing rate given a fixed synaptic input rate $\gamma$.  The firing rate $f$ is the number of spikes emitted by a neuron per unit time, $f= N_{\rm{sp}}/ T$, where $N_{\rm{sp} }$ is the number of spikes and $T$ is the observation time interval.
To calculate the {\it f-I} curve, we simulated Eqs (\ref{eq:LI_eq}), (\ref{eq:Dyn_th}), and (\ref{eq:I_syn}) with time step $\delta t$= 0.01[ms] and the time interval $T$= 50[s].

%%%%%%%%%%%%%%%%%%%%%%%%%%%%%%
%%% Results
%%%%%%%%%%%%%%%%%%%%%%%%%%%%%%
\section{Results}
\subsection{Irreproducibility of the f-I curve by the LIF neuron}
We compared the {\it f-I} curves of a cortical neuron obtained by Chance et al.~\cite{Chance}  (Figure 1A) to those of the LIF neuron (Figure 1B). 
We found that the LIF neuron cannot reproduce two main features of the {\it f-I} curves of a cortical neuron.  The first is the onset of the {\it f-I} curve. The onset of the {\it f-I} curves of a cortical neuron are linear, whereas those of the LIF neuron are nonlinear.  The second irreproducible feature is the asymptotic behavior of the {\it f-I} curve.  The asymptotic behavior of the {\it f-I} curves of a cortical neuron are sublinear, whereas those of the LIF neuron are linear.\\
\begin{figure}[htb]  % Fig 1
   \begin{center}
     \includegraphics[width=12cm]{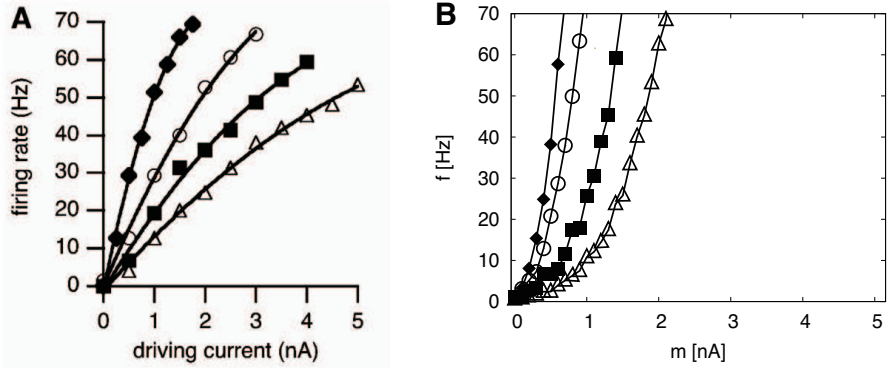}
     \caption{Comparison of the {\it f-I} curves of a cortical neuron and the LIF neuron. (A) The {\it f-I} curves of a cortical neuron with varying  levels of synaptic input rate (adapted from Chance et al.~\cite{Chance} with kind permission from the author and Elsevier). (B) The {\it f-I} curves of the LIF neuron. Synaptic input rate $\gamma= 67.5$[Hz] (closed diamonds), $\gamma= 135$[Hz] (open circles), $\gamma= 270$[Hz] (closed squares), $\gamma= 405$[Hz] (open triangles).}
     \label{fig:LIF_FI}
   \end{center}
\end{figure}
\subsection{Effect of the dynamical threshold on the {\it f-I} curve}
\begin{figure}[htb]  %  Fig 2:  
   \begin{center}     %  (A) Onset  (B) Nonlinearity
     \includegraphics[width=12cm]{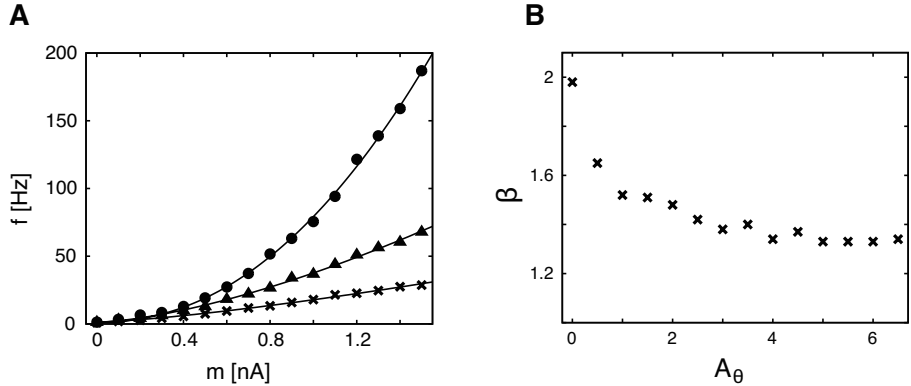}
     \caption{The effect of the dynamical threshold on the {\it f-I} curve.  (A) The {\it f-I} curves of the LIFDT neurons fitted by the power function (lines). Parameters: $\gamma= 135$[Hz] (fixed) while $A_{\theta}= 0$ (circles), $A_{\theta}= 0.5$ (triangles), $A_{\theta}=3$ (crosses).  (B) The nonlinearity $\beta$ of the {\it f-I} curve is plotted as a function of $A_{\theta}$. Parameters: $\gamma= 135$[Hz] (fixed).}
     \label{fig:FI_onset}
   \end{center}
 \end{figure}

\begin{figure}[htb]  %  Fig 3: Asymptotic Behavior.  
 \begin{center}       
     \includegraphics[width=12cm]{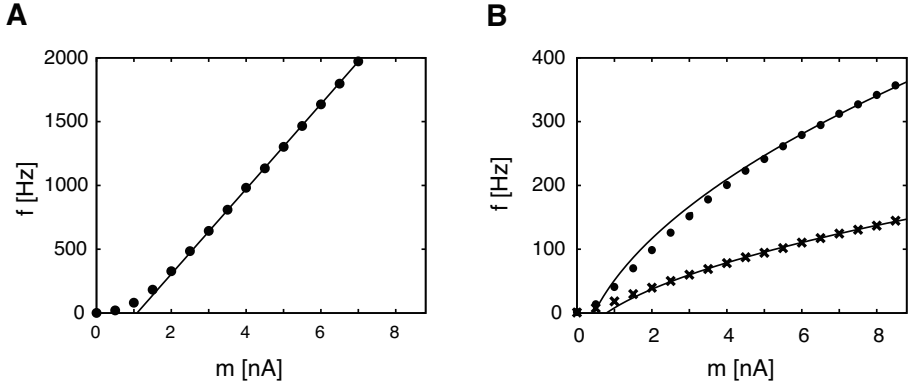}
     \caption{The effect of the dynamical threshold on the asymptotic behavior of the {\it f-I} curve. 
     (A) The {\it f-I} curve of the LIF neuron. The solid line is the fit of Eq.(\ref{eq:asympt_LIF}). Parameter: $\gamma= 135$[Hz].  (B) The {\it f-I} curve of the LIFDT neurons.  Solid lines are the fit of Eq.(\ref{eq:asympt_DT}).  Parameters: $\gamma= 135$[Hz] (fixed) while $A_{\theta}= 1.0$ (circles), $A_{\theta}= 5.0$ (crosses).}
     \label{fig:FI_asympt}
   \end{center}   
 \end{figure}

We investigated the effect of the dynamical threshold on the onset of the {\it f-I} curve.  To quantify the nonlinearity of the onset, we fitted the onset of the {\it f-I} curve by using the power function
\[ h(x)= c_1 x^{\beta}+ c_0, \]
where $\beta$ represents the nonlinearity of the onset, and $c_0$, $c_1$ are the parameters.
Figure \ref{fig:FI_onset}A shows the onset of the {\it f-I} curves of the LIFDT neuron with three values of $A_{\theta}$ while the synaptic input rate $\gamma$ is kept fixed.  
Figure \ref{fig:FI_onset}B shows the dependence of the onset nonlinearity $\beta$ on $A_{\theta}$. The onset of the LIF neuron $(A_{\theta}= 0)$ is nonlinear$-\beta= 2.0-$ while the onset of the LIFDT neuron with large $A_{\theta}$ is linear-like: $\beta= 1.3$.  The dynamical threshold linearizes the onset of the {\it f-I} curve.\\
Next, we investigated the effect of the dynamical threshold on the asymptotic behavior of the {\it f-I} curve.  Here, we derived the asymptotic formulae of the {\it f-I} curve.
We used the diffusion approximation~\cite{LaCamera2004, Lansky1999}, wherein the excitatory and inhibitory synaptic conductances are approximated by the diffusion processes, and then we neglected the time correlation of the synaptic conductances,
\begin{equation}
	g_{\scriptstyle \rm{E} }(t) \approx  a_{\scriptstyle \rm{E} } \tau_{\scriptstyle \rm{E} } \gamma+ a_{\scriptstyle \rm{E} } \sqrt{ \frac{ \tau_{\scriptstyle \rm{E} } \gamma}{2} } \xi_{\scriptstyle \rm{E} }(t),\quad   
	g_{\scriptstyle \rm{I} }(t) \approx  a_{\scriptstyle \rm{I} } \tau_{\scriptstyle \rm{I} } \gamma+ a_{\scriptstyle \rm{I} } \sqrt{ \frac{ \tau_{\scriptstyle \rm{I} } \gamma}{2} } \xi_{\scriptstyle \rm{I} } (t)  \label{eq:dif_approx}
\end{equation}
where $\xi_{\scriptstyle \rm{E},\ \rm{I} }(t)$ are independent Gaussian white-noise processes of zero mean and unit SD.
Using Eqs (\ref{eq:LI_eq}), (\ref{eq:I_syn}), and (\ref{eq:dif_approx}), we obtain
\begin{equation}
	C \frac{dV}{dt}= - g^*_{\scriptstyle \rm{L} } (V-V_{\infty}) + m+ s(V) \xi(t)
	\label{eq:LIF_syn}
\end{equation}
where
\[	g^*_{\scriptstyle \rm{L} }= g_{\scriptstyle \rm{L} }+ \gamma( a_{\scriptstyle \rm{E} } \tau_{\scriptstyle \rm{E} } + a_{\scriptstyle \rm{I} } \tau_{\scriptstyle \rm{I} }), \]
\[	s^2(V)= \frac{\gamma}{2} \{ a_{\scriptstyle \rm{E} }^2 \tau_{\scriptstyle \rm{E} } (V-V_{\scriptstyle \rm{E} })^2+ a_{\scriptstyle \rm{I} }^2 \tau_{ \scriptstyle \rm{I} } (V-V_{\scriptstyle \rm{I} })^2 \}, \]
Here, $\xi(t)$ is a Gaussian white-noise process of zero mean and unit SD, and we assume that the synaptic input is balanced,
$ ( a_{\scriptstyle \rm{E} } \tau_{ \scriptstyle \rm{E}}+ a_{\scriptstyle \rm{I} } \tau_{ \scriptstyle \rm{I}} ) V_{\infty}= a_{\scriptstyle \rm{E} } \tau_{ \scriptstyle \rm{E}} V_{ \scriptstyle \rm{E}} + a_{\scriptstyle \rm{I} } \tau_{ \scriptstyle \rm{I}} V_{\scriptstyle \rm{I} }$.
We write the firing rate of the LIF neuron (Eq. (\ref{eq:LIF_syn}) ) with the threshold $\theta_0$ as $f_{\rm LIF}(m, s(V), \theta_0 ) $, and the firing rate of the LIFDT neuron as $f_{\rm DT}(m, s(V) )$.  
For large constant input $s(V) \ll m$, we can neglect the contribution of the noise $s(V)$,
\[
	f_{\rm LIF}(m, s, \theta_0) \approx f_{\rm LIF}(m, 0, \theta_0),
	\quad f_{\rm DT}(m, s) \approx f_{\rm DT}(m, 0).
\]
The firing rate of the LIF neuron can be written as
\begin{equation}
	f_{\rm LIF}(m, 0, \theta_0)= \frac{ g^*_{\scriptstyle \rm{L} } }{C 
	 \log \left( \frac{m/ g^*_{\scriptstyle \rm{L} }+ V_{\infty}- V_{\rm r}  }{m/ g^*_{\scriptstyle \rm{L} }+ V_{\infty}- \theta_0} \right)} \sim 
	\frac{m}{ C (\theta_0-V_r) }+ const.  \label{eq:asympt_LIF}	
\end{equation}
The firing rate of the LIFDT neuron is given by the self-consistent solution~\cite{LaCamera2004} of
\begin{equation}
	f_{\rm DT}(m, 0)= f_{\rm LIF}(\ m, 0, \theta_{\infty}+ A_{\theta} \tau_{\theta} f_{\rm DT}(m, 0)\ ).\label{eq:self_consist}
\end{equation}
From Eq. (\ref{eq:self_consist}), we can obtain
\begin{equation}
	f_{\rm DT}(m, 0)= \frac{\sqrt{ (\theta_{\infty}-V_{\rm r})^2 + 4 m A_{\theta} \tau_{\theta}/ C}- (\theta_{\infty}-V_{\rm r}) }{2 A_{\theta} \tau_{\theta} }	
	\sim \sqrt{ \frac{m}{ C A_{\theta} \tau_{\theta}} }+ const. \quad
	\label{eq:asympt_DT}
\end{equation}
Thus we can obtain the asymptotic formula of the {\it f-I} curve for large $m$,
\[     f_{\rm LIF} \sim \frac{m}{ C (\theta_0-V_r) }, \quad f_{\rm DT} \sim \sqrt{ \frac{m}{ C A_{\theta} \tau_{\theta}} }. \]
Figure \ref{fig:FI_asympt}A shows the {\it f-I} curve of the LIF neuron, and Figure \ref{fig:FI_asympt}B shows the {\it f-I} curves of the LIFDT neuron.  In the large $m$, these {\it f-I} curves approach Eqs (\ref{eq:asympt_LIF}) and (\ref{eq:asympt_DT}).
The dynamical threshold changes the asymptotic behavior of the {\it f-I} curves: the asymptotic behavior of the LIF neuron is linear $(f \sim m)$, whereas that of the LIFDT neuron is sublinear $(f \sim m^{1/2})$.

\subsection{Reproducibility of the f-I curve by the LIFDT neuron}

We compared the {\it f-I} curves of a cortical neuron obtained by Chance et al.~\cite{Chance}  (Figure 4A) to those of the LIFDT neuron (Figure 4B).
The LIFDT neuron can reproduce the two main features of the {\it f-I} curve of a cortical neuron: the linear onset and the sublinear asymptotic behavior. 
  Our results are summarized in Table \ref{table:Comp_neuron}.
\begin{table}[htb]
\begin{tabular}{|c|c|c|}   \hline
	Neuron   & Onset  &  Asymptotic Behavior  \\   \hline
	Experiment~\cite{Chance}       &  Linear  &  Sublinear  \\   \hline
	LIF                       &  Nonlinear  &  Linear  ($\sim m$) \\ \hline
	LIFDT       &  Linear  &  Sublinear ($\sim m^{1/2}$)  \\ \hline	
\end{tabular}
\caption{Main features of the {\it f-I} curves}
\label{table:Comp_neuron}
\end{table}

\begin{figure}[htb]  %  Fig 4: LIFDT
   \begin{center}
     \includegraphics[width=12cm]{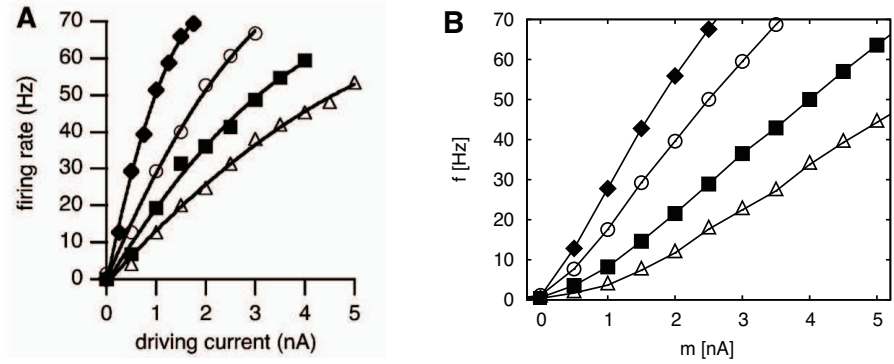}
     \caption{ Comparison of the {\it f-I} curves of a cortical neuron and a LIFDT neuron.
     (A) The {\it f-I} curves of a cortical neuron with varying  levels of synaptic input rate (adapted from Chance et al.~\cite{Chance} with kind permission from the author and Elsevier).    
     (B) The {\it f-I} curves of a LIFDT neuron. Parameters: $A_{\theta}= 5.0$ (fixed), synaptic input rate $\gamma= 67.5$[Hz] (closed diamonds), $\gamma= 135$[Hz] (open circles), $\gamma= 270$[Hz] (closed squares), $\gamma= 405$[Hz] (open triangles).}
     \label{fig:FI_LIFDT}
   \end{center}
 \end{figure}
% \newpage

%%%%%%%%%%%%%%%%%%%%%%%%%%%%%%
%%% Discussion
%%%%%%%%%%%%%%%%%%%%%%%%%%%%%%
\section{Discussion}
\subsection{Conclusion}    
We studied the effect of a dynamical threshold on the {\it f-I} curve. 
It is confirmed that the dynamical threshold is essential for reproducing the {\it f-I} curve of a cortical neuron and that it modulates the onset and the asymptotic behavior of the {\it f-I} curve. These results suggest that a cortical neuron has an adaptation mechanism and that
this significantly influences the computation of a cortical neuron.

\subsection{Unresolved problems}
Here we state two unresolved problems relating to our study.  
The first is the clarification of the effect of adaptation on the network behavior of neurons.  It would be interesting to study the effect of adaptation on not only the behavior of a neuron but also on that of interacting neurons.
The second open problem is the clarification of the effect of the adaptation mechanism on the Hodgkin$-$Huxley-type model neurons~\cite{Chacron2007}.  For the sake of simplicity, we studied LIF and LIFDT neurons in this paper.   However, Hodgkin$-$Huxley-type models are known to be more realistic neuron models.  It is currently thought that the adaptation mechanism mainly arises from M-type currents, mAHP-type currents, and slow sodium currents in the Hodgkin$-$Huxley-type model~\cite{Benda2003, Ermentrout1998, Koch1999}.  It would be interesting to study the effect of these currents on the {\it f-I} curve.

%%%%%%%%%%%%%%%%%%%%%%%%%%%%%%
%%% Acknowledgments
%%%%%%%%%%%%%%%%%%%%%%%%%%%%%%
\ack{
We thank the organizing committee of UPoN 2008 for an interesting conference.  We thank Frances Chance and Elsevier Limited to use figures from Chance et al.~\cite{Chance}.  We also thank Shigeru Shinomoto for fruitful discussions, and Kensuke Arai and Shigefumi Hata for useful comments on the manuscript.
This study is supported by a Grant-in-Aid for JSPS Fellows from the Japan Society for the Promotion of Science.  R.K. is supported by the Research Fellowship of the Japan Society for the Promotion of Science for Young Scientists.}

%%%%%%%%%%%%%%%%%%%%%%%%%%%%%%
%%% References
%%%%%%%%%%%%%%%%%%%%%%%%%%%%%%

\end{document}